\documentclass[]{interact}
\usepackage{epstopdf}
\usepackage{xcolor}

\usepackage[natbibapa,nodoi]{apacite}
\theoremstyle{plain}

\theoremstyle{definition}

\theoremstyle{remark}

\begin{document}

\articletype{Review Article}

\title{When Generative Artificial Intelligence meets Extended Reality: A Systematic Review}

\author{
\name{Xinyu, NING\textsuperscript{a}, Yan ZHUO\textsuperscript{a}, Xian WANG\textsuperscript{a}, Chan-In Devin SIO\textsuperscript{a}, and Lik-Hang LEE\textsuperscript{a}\thanks{The corresponding author: Lik-Hang LEE, Email: lhleeac@connect.ust.hk; Ning and Zhuo equally contribute to this work (as co-first authors).}}
\affil{\textsuperscript{a}The Hong Kong Polytechnic University, Department of Industrial and Systems Engineering, Hong Kong SAR, China}
}

\maketitle

\begin{abstract}
With the continuous advancement of technology, the application of generative artificial intelligence (AI) in various fields is gradually demonstrating great potential, particularly when combined with Extended Reality (XR), creating unprecedented possibilities. This survey article systematically reviews the applications of generative AI in XR, covering as much relevant literature as possible from 2023 to 2025. The application areas of generative AI in XR and its key technology implementations are summarised through PRISMA screening and analysis of \textcolor{black}{the final 26 articles.} The survey highlights existing articles from the last three years related to how XR utilises generative AI, providing insights into current trends and research gaps. We also explore potential opportunities for future research to \textcolor{black}{further empower} XR through generative AI, providing guidance and information for future generative XR research.  
\end{abstract}




\section{Introduction}

Generative artificial intelligence (AI) and Extended Reality (XR) are among the most revolutionary innovations in today's technologies (\citeauthor{[1]}, \citeyear{[1]}). Generative artificial intelligence can create entirely new data, images, audio, or video content based on a given input through techniques such as machine learning and deep learning, thereby significantly broadening the boundaries of digital creation (\citeauthor{[2]}, \citeyear{[2]}). 
XR technology, encompassing augmented reality (AR), virtual reality (VR), and mixed reality (MR), offers users an immersive and interactive experience that seamlessly blends the virtual and real worlds (\citeauthor{[3]}, \citeyear{[3]}). With the increasing intersection of the two, the application of generative AI in XR is beginning to show tremendous potential, particularly in fields such as gaming, education, healthcare, and industry, among others, driving the emergence of numerous innovative applications (\citeauthor{[4]}, \citeyear{[4]}; \citeauthor{[5]}, \citeyear{[5]}).
However, despite the great potential of combining generative AI and XR technology, relevant research is still in its preliminary stage, and there has been a lack of systematic discussion in the \textcolor{black}{research} community on this field for the past few years. Most existing research focuses on a single application scenario or technical level and lacks a comprehensive review and analysis. To address this gap, this \textcolor{black}{article} employs the PRISMA (Preferred Reporting Items for Systematic Reviews and Meta-Analyses) method to systematically review the application of generative artificial intelligence in XR (\citeauthor{[6]}, \citeyear{[6]}). The aim is to summarise the current research results, identify the main research trends, and provide references for future research.

Specifically, the main objective of this article is to offer a \textcolor{black}{concise yet systematic} review of emergent generative artificial intelligence applications in the field of XR. \textcolor{black}{The majority of these works are collected from the databases of computer science and engineering (e.g., ACM \& IEEE). Our statistics facilitate a discourse on their practical uses and possibilities across many sectors.}  
Combining the existing literature, we identify different application scenarios generated by \textcolor{black}{artificial intelligence and XR technology, as well as the challenges and opportunities associated with pattern design. }
A future research direction \textcolor{black}{accordingly} is proposed, aiming to address the shortcomings of the selected scope from the existing literature and therefore identify unexplored areas of research \textcolor{black}{related to human-computer interaction, e.g., security and privacy, user autonomy, and child protection. We outline the three key findings of our study and the corresponding future directions, as follows. \\
\begin{enumerate}
    \item Contemporary generative AI technologies primarily concentrate on producing text, images, and various graphical assets, thereby facilitating the creation and instantiation of objects within virtual environments. There is insufficient consideration for other input or output modalities, including more complex user behaviours. 
    \item The majority of the existing studies attempt to demonstrate the potential XR applications enabled by diverse generative AI models, e.g, education and training for content creation. Nonetheless, the connection between generative AI and the underlying principle of human-computer interfaces is often neglected. 
    \item Generative AI primarily functions as an enabler in XR, delivering unimodal output based on a snapshot of user inputs. We advocate additional research efforts for investigating how the future role of generative AI will transcend its existing capabilities, acting as a vital integrator of various modalities in XR and as a dedicated agent of user experiences in the forthcoming research agenda of human-computer interaction.
\end{enumerate}
}

The structure of this article is as follows. First, we introduce detailed PRISMA methods and the selection process of the literature. Then, we present the application status of generative artificial intelligence in XR, and analyse related technologies and cases; Afterwards, we summarise the main findings in the literature, and discuss the current application challenges and opportunities; Finally, our article puts forward some suggestions for future research. 

\section{Methodology}
\begin{figure}[t]
\centering
\includegraphics[width=0.9\columnwidth]{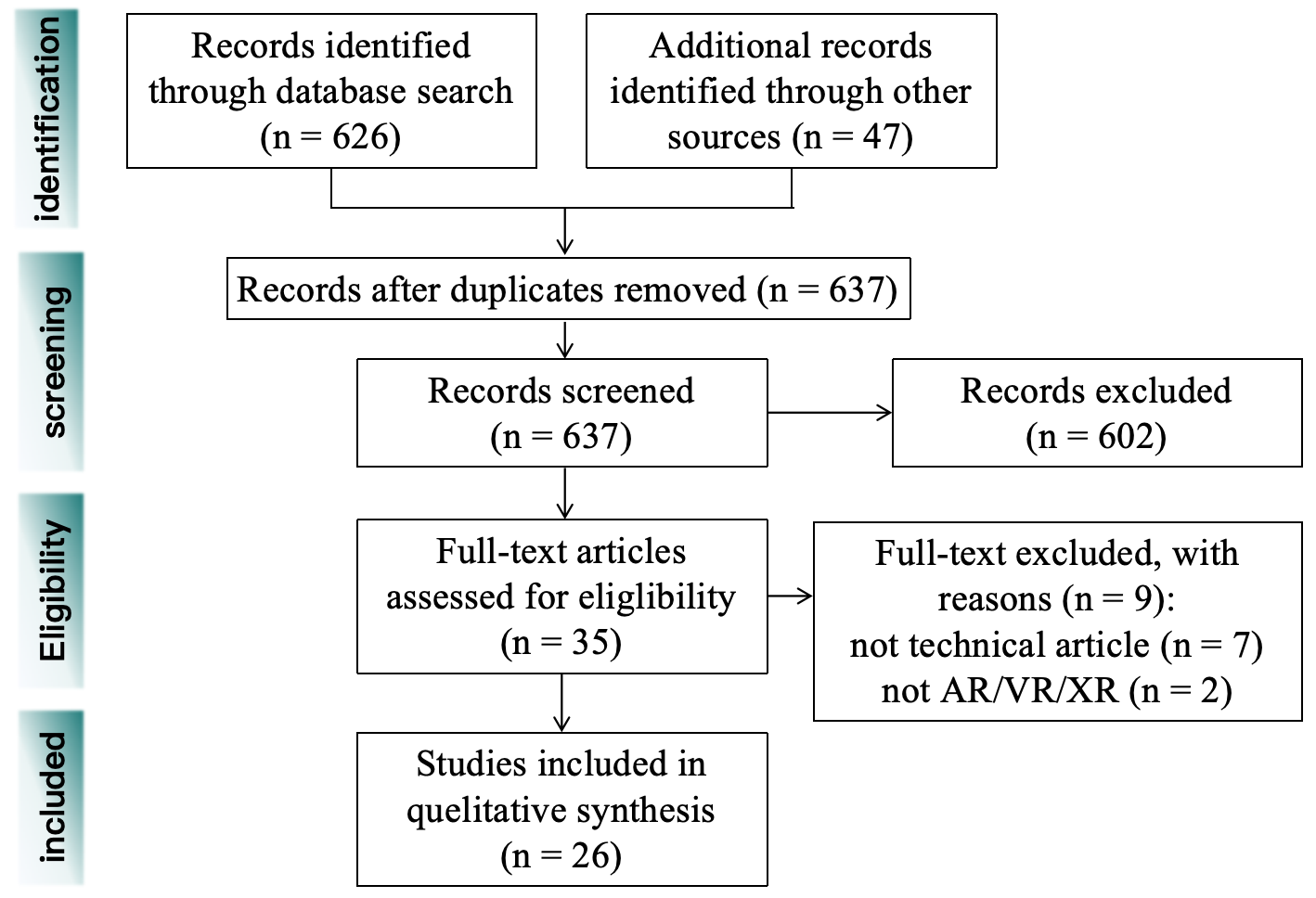}
\caption{PRISMA flow diagram for systematic review.}
\label{fig:f1}
\end{figure}

To capture the most recent studies, our literature review employed a two-stage approach, with the first data extraction conducted in February 2025 and the second in April 2025. Complied with the systematic approach, namely PRISMA (Preferred Reporting Items for Systematic Reviews and Meta-Analyses), we selected 26 articles \textcolor{black}{(see the $*$ symbol and [number] (refer to Figure 7) at the title of each selected entry in the reference list)} for data extraction and further analysis.
PRISMA is a set of standard guidelines for conducting systematic literature reviews and meta-analyses (\citeauthor{[7]}, \citeyear{[7]}). The system can comprehensively, clearly, and thoroughly explain the research methods and research results in a particular field, and then accurately evaluate the reliability and practical application value of research in this field. To ensure the rigour and transparency of our literature review process, the review was conducted in collaboration with the (co-)first authors to proceed with a systematic review (\citeauthor{[8]}, \citeyear{[8]}) and meta-analysis (\citeauthor{[9]}, \citeyear{[9]}). Figure \ref{fig:f1} depicts the complete PRISMA procedures and the key results. After completing the initial search, we identified 673 candidate articles. Then, we initiated a structu filter protocol, which resulted in the removal of 36 duplicate articles, leaving 637 articles for the subsequent screening process. We screened these articles by title and abstract, excluding 602 articles that did not meet our criteria. After full-text evaluation, a further nine articles were excluded. The specific inclusion and exclusion criteria are detailed in the following paragraphs.

\subsection{Search Strategies}

\subsubsection{Keywords}
Keywords mainly reflect the application of the generative AI model, content creation, and XR technology in three dimensions. Taking into account the terms expressed in different contexts, we chose the following keywords to describe the XR environment supported by generative AI: “generative AI”, “intelligent”, “content generation”, “algorithm-driven”, “automated authoring”, “virtual synthesis”, and “digital world building”. 
Other essential keywords related to generative AI models include “GANs” (generative adversarial networks), “diffusion model”, “converter architecture” (such as transformer), “data-driven design”, “image generation”, “3D scene generation”, “automatic content generation”, “personalized content generation” and “real-time content update”, “depth learning”, “neural networks”. 
Keywords and acronyms related to XR technology include “virtual reality”, “augmented reality”, “mixed reality”, “VR”, “AR”, and “MR” (\citeauthor{[33]}, \citeyear{[33]}) \textcolor{black}{While the primary focus of this article differs from the studies focusing on manufacturing and business sectors, it is worth noting that recent research has investigated the applications of generative artificial intelligence in extended reality across various domains, including marketing analytics, industrial digital twins, and enterprise management within the metaverse (\citeauthor{[75]}, \citeyear{[75]}; \citeauthor{[76]}, \citeyear{[76]}; \citeauthor{[77]}, \citeyear{[77]}). These works provide valuable insights into the broader landscape of generative AI and XR technologies, which may inform future interdisciplinary developments. Nonetheless, we did not include them in the search keywords due to our research objective. }

\subsubsection{Database}
To ensure the breadth and depth of the literature review, we conducted systematic paper searches in well-known databases, including IEEE Xplore, ACM Digital Library, and Elsevier’s ScienceDirect. These databases allow us to obtain sufficient papers that represent studies from diversified domains, e.g., computer science, engineering, and the related applications that bring impact cases to the real world. To expand our sources further, we utilised Google Scholar as an additional tool. We employed the snowball search method, which begins with the initial selection of relevant literature and tracks citations to identify more valuable references. This method not only helps us find more directly relevant research but also enables us to establish a more comprehensive research background framework. 
\textcolor{black}{We acknowledge the limitations of platforms like Google Scholar, particularly regarding quality filtering. To address this, we applied rigorous manual screening during article selection, focusing on relevance, methodological soundness, and citation impact where available with the PRISMA approach. While we did not employ automated methodological quality assessment tools such as Abstrackr or ROBIS, our review process involved independent evaluation by multiple authors to ensure the inclusion of high-quality and pertinent literature, primarily from the well-known computer science and engineering databases. We believe this balanced approach allows us to capture both established and emerging research, providing a decent snapshot of the current landscape. Additionally, the inclusion of works from ArXiv is to reflect the fast-changing landscape of computer science, especially for the convergence of generative AI and XR.}

\subsection{Inclusion and Exclusion Criteria}
During the process of screening and evaluating articles for compliance with the PRISMA review criteria, we established specific inclusion and exclusion criteria to aid in selecting suitable papers. Table \ref{tab:my-table1} depicts these rules and illustrates the basis by which we judge the suitability of the study. We focused primarily on studies that met at least one of the inclusion criteria, while ensuring that any articles that did not meet the criteria (i.e., those that touched any of the exclusion criteria) were excluded. As such, we pursued a comprehensive and meticulous study of the current research to guarantee that the ultimately chosen literature was both precise and representative. 

\begin{table}[]
\caption{Inclusion and Exclusion Criteria}
\label{tab:my-table1}
\begin{tabular}{|l|l|l|}
\hline
\textbf{Sort}         & \textbf{Inclusion criteria}                                                                                            & \textbf{Exclusion criteria}                                                                        \\ \hline
Research type         & Original research paper                                                                                                & Review                                                                                             \\ \hline
Technical category    & \begin{tabular}[c]{@{}l@{}}Explicit use of generative \\ AI technologies   \\ (GAN/VAE/Transformer, etc.)\end{tabular} & Non-generative AI                                                                                  \\ \hline
Application scenarios & \begin{tabular}[c]{@{}l@{}}Involving XR terminals   \\ (VR/AR/MR Devices)\end{tabular}                                 & \begin{tabular}[c]{@{}l@{}}Only theoretical discussion  \\ without empirical research\end{tabular} \\ \hline
Literature language   & English literature                                                                                                     & Non-English literature                                                                             \\ \hline
\end{tabular}
\end{table}

\subsection{Data Extraction}
To extract relevant information from the included articles, we developed a data extraction standard that systematically captures key aspects of the application of augmented, virtual, or mixed reality technologies and AI-generated models. Initially, we randomly selected five articles and developed the project based on the relevant aspects identified within them. The initial data extraction criteria were revised to the final version after evaluation by both authors, as shown in Table \ref{tab:my-table2} (\citeauthor{[33]}, \citeyear{[33]}). The data for each article is extracted independently according to this final standard. In the case of data conflict, the two authors reach a consensus through discussion.

During the data extraction phase, we meticulously recorded the basic information of each article, including the study number (author name, publication date), article title, and keywords. For the application of XR technology, we focus on standard technologies, including augmented reality (AR), virtual reality (VR), and mixed reality (MR), which involve devices such as VR headsets, AR glasses, and mobile AR. In particular, the various ways in which different hardware can be used in virtual environments are highlighted. XR system can be enhanced by adding features such as auditory feedback, video overlay, 3D image layers, avatars, and more, and can be deployed or upgraded in multiple locations. We also focus on research evaluation methods, which can be quantitative analysis, qualitative analysis based on questionnaires, or a discussion of latency.

\begin{table}[]
\caption{Data Extraction Rubric for the Selected 26 Articles}
\label{tab:my-table2}
\begin{tabular}{|l|l|}
\hline
\textbf{Data extraction dimension} & \textbf{Content}                                                \\ \hline
Study number                       & Source, date of publication                                     \\ \hline
Title                              & Open text                                                       \\ \hline
Keywords                           & Open text                                                       \\ \hline
XR technology used                 & VR, AR, MR, others                                                 \\ \hline
XR devices used                    & HMD, Mobile                                                      \\ \hline
Technical models                   & GAN, VLM, LLM, Diffusion, Transformer                               \\ \hline
User interaction modes             & Controller, gesture, natural voice input, image, eye   tracking \\ \hline
Enhanced information type          & Graphics, text, speech, action                                  \\ \hline
Experimental design                & Sample size/Evaluation indicators                               \\ \hline
Innovation contribution            & Performance improvement                                         \\ \hline
Limitation                         & Data dependency/Data latency                                    \\ \hline
\end{tabular}
\end{table}

\section{RESULTS}

\subsection{Overview of Included Articles}
We carefully extracted relevant information from the 26 articles screened. The selected articles span from 2023 to 2025 (Q1), and the number of articles published per year is shown in Figure 2. These articles are from well-known XR venues. 
The conferences include: the IEEE International Conference on Metaverse Computing, IEEE International Conference on Artificial Intelligence and eXtended and Virtual Reality (AIxVR)conference, IEEE International Conference on Metaverse Computing, Networking, and Applications (MetaCom), Proceedings of the 2023 ACM Symposium on Spatial User Interaction conference, ACM Symposium on Virtual Reality Software and Technology, etc. 
On the other hand, journals include: Computers and Education: Artificial Intelligence, Computers \& Education: X Reality, and others. These data were then analysed and summarised, with additional qualitative insights gained during the iterative analysis process. Over the past decade, there has been a general increase in the number of articles on the topics investigated, suggesting that interest in and emphasis on these topics have grown in recent years. Among the 26 articles included, the highest number of publications was observed in 2024 (\citeauthor{[10]}, \citeyear{[10]}). Given that the literature from 2025 only covers the first half of the year, the trend observed in the collected studies suggests that future research will increasingly focus on the convergence of generative AI with XR technologies.

\begin{figure}[t]
\centering
\includegraphics[width=0.9\columnwidth]{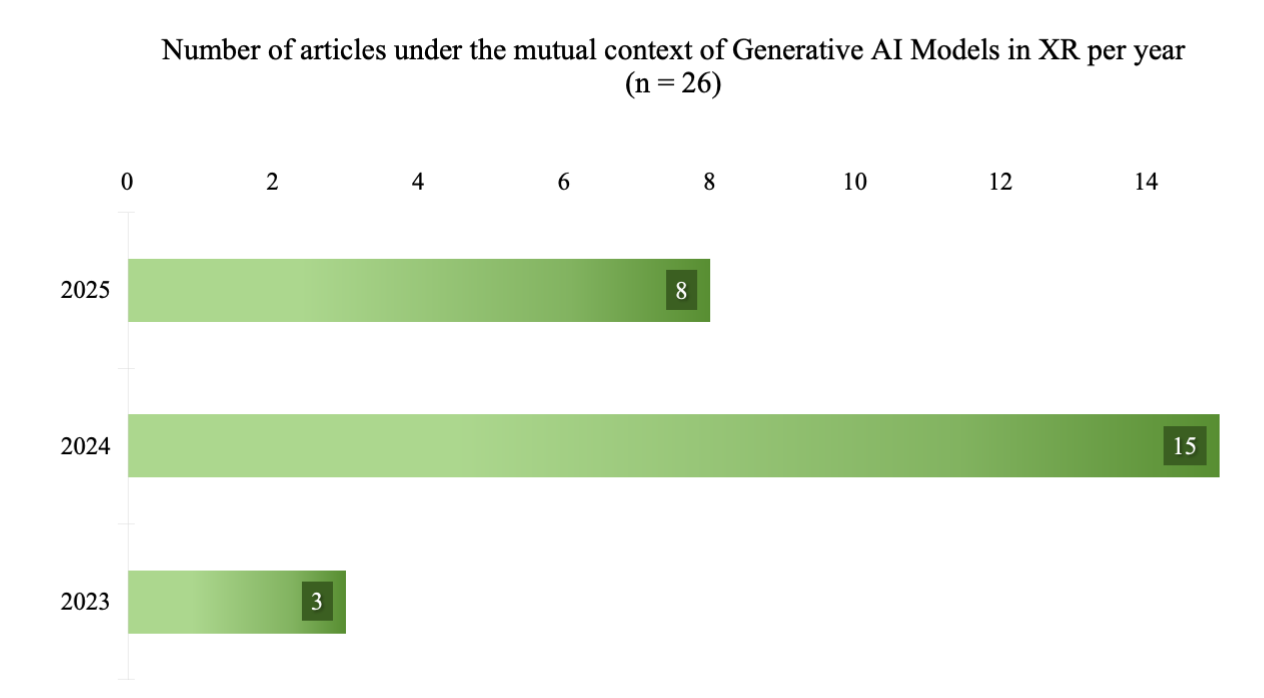}
\caption{Selected articles per year (N=26). We observe a huge surge in 2024. The collected article in 2025 is less than in 2024 because our search period stopped at Q1 2025.}
\label{fig:f2}
\end{figure}

\subsection{Keywords Analysis}
By analysing the keywords in the literature, we find that the terms “reality” and “virtual” appear frequently, with the term “virtual” being significantly more frequent than “augmented”. The frequency of the word “virtual” is significantly higher than that of “augmented”, indicating that virtual reality (VR) is more widely used than augmented reality (AR) in the current research. Meanwhile, the frequency of the term “extended reality” (XR) as a collective term for MR, AR, and VR is lower than that of MR, indicating that its application scope is not broad enough and is generalised in many current research methods.
In addition, “generative AI”, “XR”, and content generation-related technologies (e.g., text-to-image, 3D generation, etc.) are also high-frequency keywords. This indicates that research in this field is gradually shifting to how to utilize generative AI to enhance the content generation capability of XR technology. Figure 3 shows the keywords of 26 articles in the form of word clouds, thus showing more accurately the frequency distribution of these words in the included articles.

\begin{figure}[t]
\centering
\includegraphics[width=0.9\columnwidth]{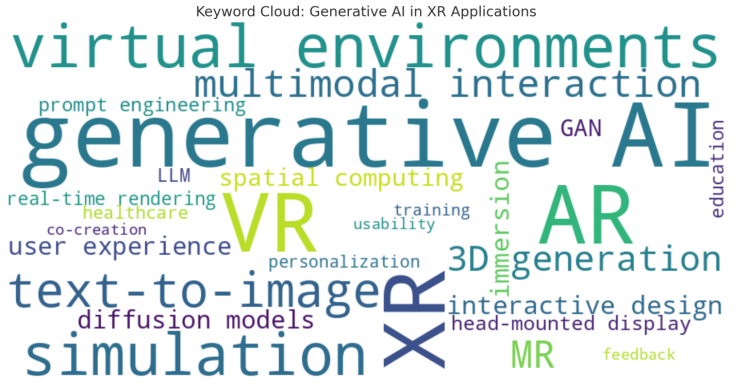}
\caption{Frequently-used keywords in the included articles.}
\label{fig:f3}
\end{figure}

\subsection{Application Domains}
\textcolor{black}{In our analysis of 26 included papers, the application domains of generative AI combined with Extended Reality (XR) cover a wide range of fields. We classified the application domains of this literature into 9 categories: design, education and training, social interaction, medical therapy, industry, architecture, transportation, cultural heritage, and entertainment. Figure 4 presents the statistical results of the 26 pieces of literature across the 9 application domains in the form of bar charts. These categorisation results will be analysed in detail below. It is necessary to note that the total number of applications across different domains exceeds 26, as some studies span multiple application areas.}

\begin{figure}[t]
\centering
\includegraphics[width=0.9\columnwidth]{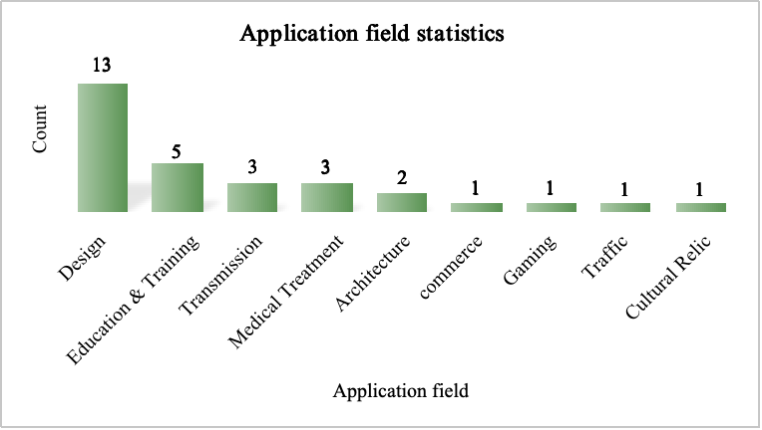}
\caption{Application domains in the included articles.}
\label{fig:f4}
\end{figure}

\subsubsection{Design and Education Training}
\textcolor{black}{Design, as well as education and training, are the two most widely used fields in the literature, accounting for 43\% and 17\% of the collected articles, respectively. According to the statistical counts, the combination of generative AI and XR technology has been applied primarily in the design field, mainly in areas such as virtual prototyping, architectural design, and interactive product design (\citeauthor{[34]}, \citeyear{[34]}) For example, MS2Mesh-XR, shown in Figure 5, is a multi-modal sketch-to-mesh generation process that allows users to quickly generate high-quality 3D models through hand-drawn sketches and voice input in the XR environment (\citeauthor{[17]}, \citeyear{[17]}).}

In the field of education and training, the use of XR enables users to access more diverse information and enhance visual perception, thereby promoting participation in learning activities (\citeauthor{[35]}, \citeyear{[35]}). Therefore, using XR in education can enhance the real-world view by adding sound, video, and graphics to the learning environment, enabling learners to interact with the content (\citeauthor{[36]}, \citeyear{[36]}). The use of virtual and augmented reality devices (e.g., Google Glass or Oculus Rift) combined with generative AI can provide users with more immersive and interactive learning experiences, such as virtual experiments, simulated training, and customised learning content (\citeauthor{[37]}, \citeyear{[37]}). Figure 6 illustrates a VR environment featuring a generative AI-powered virtual assistant to support participants in answering anatomical questions of varying cognitive complexity and facilitating verbal communication (\citeauthor{[25]}, \citeyear{[25]}).

\begin{figure}[t]
\centering
\includegraphics[width=0.9\columnwidth]{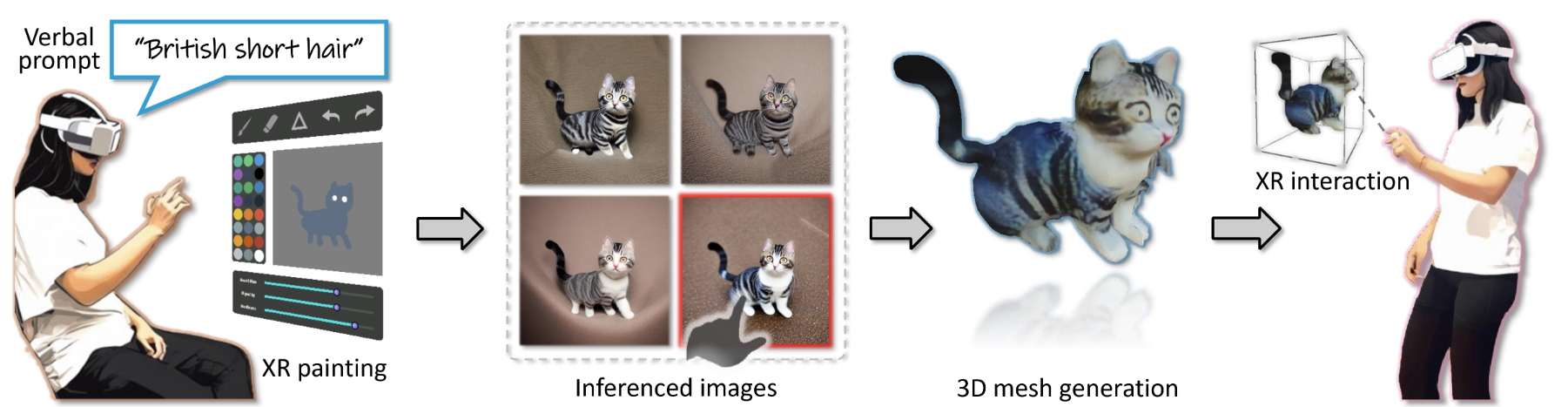}
\caption{Generating 3D models by MS2Mesh-XR: It is important to note that users with simple gestures and voice commands can generate 3D models in immersive environments.}
\label{fig:f5}
\end{figure}

\subsubsection{Transmission, Medical Treatment, and Architecture}
Transmission, medical treatment, and architecture are another group of more important application areas, covered by 10\%, 10\%, and approximately 7\% of the literature, respectively. In the transmission domain, by leveraging technologies like 5G, distributed computing, and AI, solutions are emerging that enable more efficient and personalized transmission of immersive XR content. These innovations are crucial for overcoming the challenges of real-time image capture and enhancing the overall efficiency of volumetric data transfer in XR environments (\citeauthor{[31]}, \citeyear{[31]}). In the field of medical treatment, the combination of generative AI and XR is gradually expanding its application to virtual treatment and patient health monitoring, providing more diversified treatment and rehabilitation methods (\citeauthor{[28]}, \citeyear{[28]}). In the architecture field, generative AI technology enables researchers to generate monocular 360-degree images that can be transferred to an XR device, provides a diverse range of rendering styles, enhances the immersion of architectural design and planning, and provides an innovative visualization tool for the seamless integration of architecture with the physical environment (\citeauthor{[16]}, \citeyear{[16]}).
\begin{figure}[t]
\centering
\includegraphics[width=0.9\columnwidth]{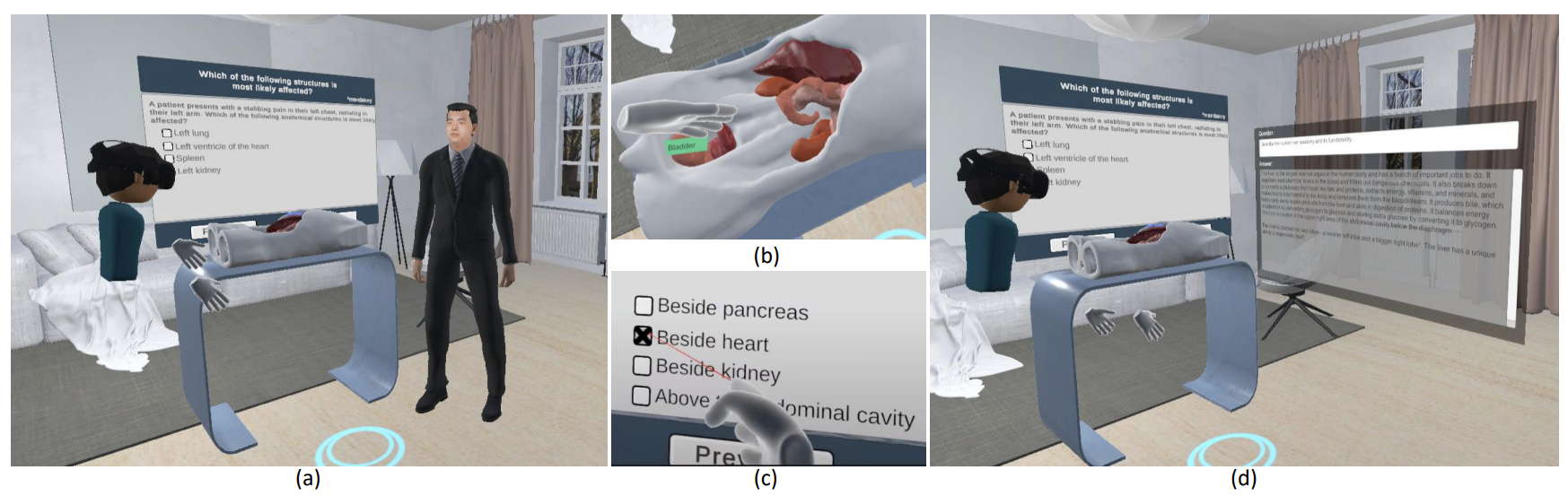}
\caption{Virtual dissolving education assistant based on generative artificial intelligence.}
\label{fig:f6}
\end{figure}

\subsubsection{Other Domains}
The literature in the fields of commerce, gaming, traffic, and cultural relics remains limited, with only one relevant study in each area. In the field of commerce, the integration of generative AI and augmented reality (AR) aims to simplify 3D model creation and manipulation, offering a system that allows users to interact with 3D models in real time without requiring specialized skills or software for e-commerce (\citeauthor{[74]}, \citeyear{[74]}). In the field of the gaming sector, the use of generative AI and virtual reality (VR) technologies has led to the development of immersive, game-based learning platforms, such as "LearningverseVR," which enhances interactive learning through AI-driven non-playable characters (NPCs) and VR environments (\citeauthor{[24]}, \citeyear{[24]}). In the field of traffic, XR technology-assisted driving simulation and traffic flow analysis are also becoming a research hot spot (\citeauthor{[21]}, \citeyear{[21]}). There is only one piece of literature in the cultural heritage domain that reflects the initial application of generative AI combined with XR in these domains (\citeauthor{[14]}, \citeyear{[14]}), which demonstrates research prospects and calls for additional research efforts. It means that XR technology and generative AI are used to reconstruct virtual monuments and sites for digital preservation and display (\citeauthor{[43]}, \citeyear{[43]}).

\begin{figure}[t]
\centering
\includegraphics[width=1\columnwidth]{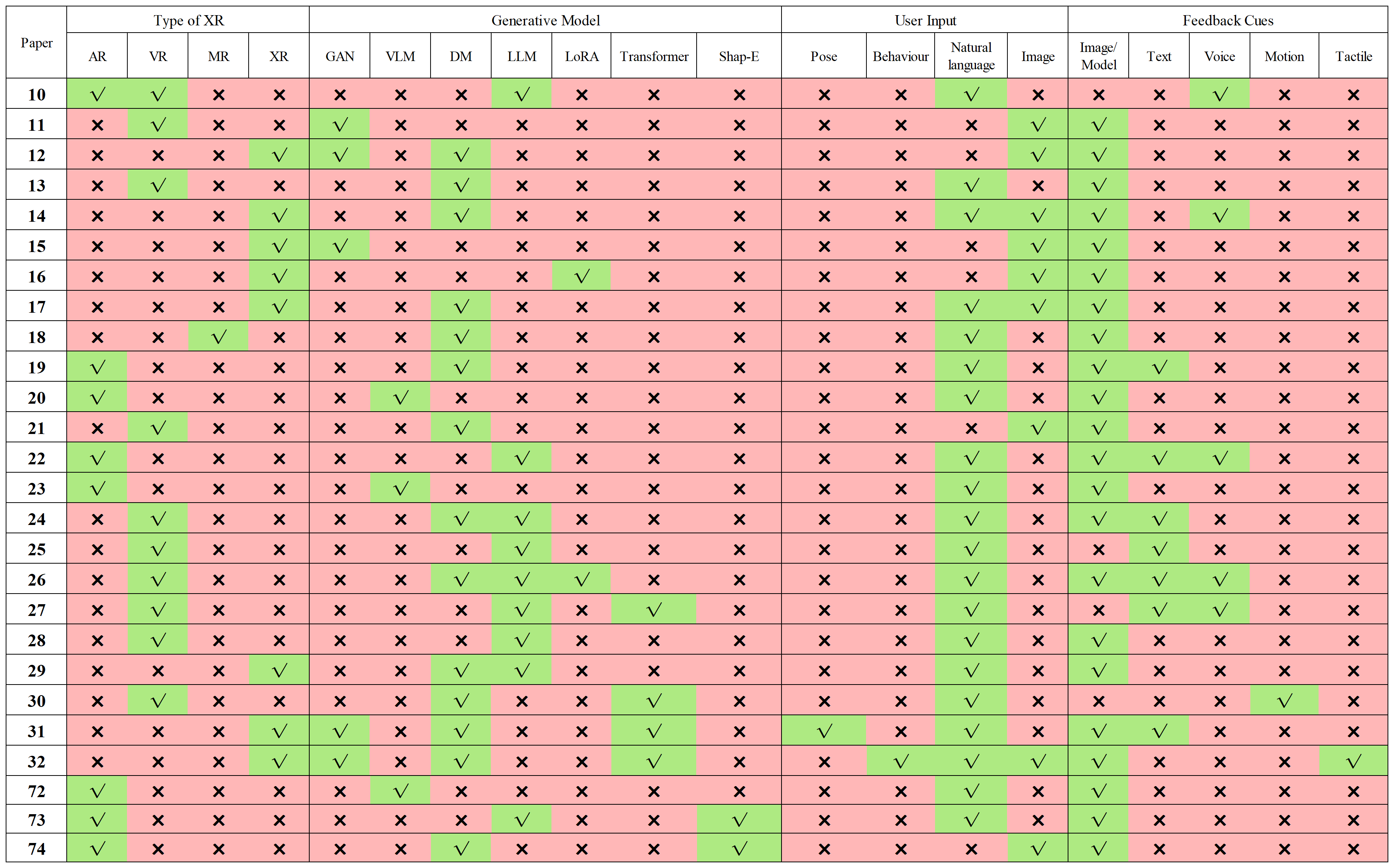}
\caption{The analysis of the combination of generative AI with Extended Reality (XR) technology, under four key dimensions: Type of XR, Generative Model, User Input, and Feedback Cue.}
\label{fig:f7}
\end{figure}

\subsection{Technology}
In the previous paragraphs, we examined the integration of generative AI and XR techniques across various domains. The following paragraphs, based on the data in Figure 7, focused on the use of different XR techniques (Virtual Reality (VR), Augmented Reality (AR), Mixed Reality (MR) as well as various types of generative AI models (e.g., Generative Adversarial Networks, Diffusion Models, Large Language Model, etc.) in the related research.

In terms of which type of XR technology they empower, Virtual Reality (VR) and Augmented Reality (AR) dominate the applications of generative AI. Specifically, Virtual Reality (VR) was applied in 10 papers, Augmented Reality (AR) was employed in 8 papers, and Mixed Reality (MR) was relatively rare, appearing in only one study. Additionally, 8 pieces of literature did not specifically differentiate between AR, VR, and MR types. Instead, they used the formulation of pan-XR or hybrid XR environments, indicating that some studies focused more on integrating cross-platform and cross-media technology and content generation (\citeauthor{[31]}, \citeyear{[31]}; \citeauthor{[32]}, \citeyear{[32]}). Therefore, from the perspective of current technology distribution, VR and AR remain the most significant landing scenarios for generative AI in XR. At the same time, MR and XR represent a more convergent and forward-looking research direction. This distribution not only reflects the difference in technology maturity but also provides a direction guide for an obvious research agenda: on the one hand, the on-going debate continue to deepen the application scenarios in VR and AR, on the other hand, the research community has been paying attention to how to promote the convergence and evolution of different XR technologies through AI (\citeauthor{[44]}, \citeyear{[44]}).

In terms of the use of generative models, a variety of generative models are involved in the study, among which Generative Adversarial Networks (GANs) appear in 5 papers, showing their potential for application in generating realistic content. Large-scale language models (LLMs) appeared in 9 articles, indicating their increasing role in generative dialogue and narrative generation. The Visual-Linguistic Model (VLM) was, however, used in three studies, reflecting the trend of combining visual and linguistic information, which provides a new direction for enhancing interactivity and immersion in XR environments. The Diffusion Model (DM) has been applied in 14 types of literature, indicating that its potential in generative tasks is being gradually explored. The Transformer Model has appeared in 4 pieces of literature among generative models, indicating that it also has potential for application when dealing with linguistic and visual tasks. Shap-E, on the other hand, was mentioned in 2 papers, highlighting its relevance in generative tasks related to shape 3D models in XR environments.

In terms of user input modes, the research explores pose input, behaviour input, natural language input, and image input as the primary interaction modes. Specifically, pose input appeared in 1 paper, and behaviour input appeared in 1 paper, indicating that these methods are still in the exploratory stage for interactive environments. Natural language input, on the other hand, is mentioned in 20 papers, both for voice input and text input, which suggests that natural language is crucial for generative AI interactions in virtual worlds. Image input interactions were discussed in 9 papers, highlighting that the use of image input is gaining interest in XR environments.

In terms of feedback cues, image or model generation was the most widely covered aspect in the research, with only four papers not mentioning it, suggesting that generating images and models holds a central position in XR applications. Text generation has been explored in 7 studies, suggesting that narrative generation and information output play a crucial role in enhancing the immersion and interactivity of XR environments. Voice generation is applied in five pieces of literature, highlighting the role of speech content generation in virtual environments, particularly in virtual assistants and dialogue systems. Motion and tactile generation are relatively rare, appearing in only 1 piece of literature, indicating that the application of motion and tactile generation in XR is still in the exploratory stage, but also has some research potential.

As a result, virtual reality and augmented reality technologies dominate research on generative AI, particularly in the application of extended modelling, demonstrating strong capabilities. Utilising natural language input is currently the most dominant interaction method. At the same time, image and model generation is the most common type of empowered content, and the applications of text and speech generation are expanding. These trends suggest that the combination of generative AI and XR technologies is driving more natural and immersive virtual experiences, and that multi-modal interaction and content generation will play an increasingly important role in future XR applications.

\section{DISCUSSIONS}
In this section, we analyse and discuss in detail the application of Generative AI in Extended Reality (XR) based on the literature data extracted by the PRISMA method. Based on the results of our extracted data, the distribution of the number of literature in each dimension covered in this review is clearly summarised. Based on the aforementioned results, this chapter will further explore the key role of generative AI in XR systems. Specifically, we examine how various types of models, which differ significantly in their interaction design and content enhancement approaches, have impacted the user experience and technical implementation of the system (\citeauthor{[52]}, \citeyear{[52]}; \citeauthor{[44]}, \citeyear{[44]}). By examining the major trends, commonalities, differences, and relevant gaps that emerge from current research, we aim to reveal the potential benefits and challenges of combining generative AI and XR technologies and provide insights for future research directions (\citeauthor{[45]}, \citeyear{[45]}). 


\subsection{Impact of Different Types of Generative AI Models in XR Applications} 
When integrating generative AI into an XR system, the type of model chosen significantly influences aspects such as user interaction, system architecture, and the presentation of augmented content (\citeauthor{[46]}, \citeyear{[46]}). In the following, we explore the specific impact of different model types (Generative Adversarial Networks (GANs), Visual Language Models (VLMs), Image Generation Models (e.g., Stable Diffusion), Large Language Models (LLMs), and generalized transformer architectures) on the design of an XR system from two perspectives.
Different types of generative models have varying forms of data inputs and outputs, and thus impose distinct requirements on the design of the interaction logic of XR systems (\citeauthor{[46]}, \citeyear{[46]}). In the current framework of generative AI development, various types of models have emerged, exhibiting significant differences in terms of generation type and application scenarios (\citeauthor{[47]}, \citeyear{[47]}). Transformer architecture is the underlying framework of many modern natural language processing models and serves as the basis for building LLMS and VLMS (\citeauthor{[49]}, \citeyear{[49]};\citeauthor{[48]}, \citeyear{[48]}). This model specialises in processing complex sequential data and supports multi-modal information processing. VLM fuses image and text understanding and can generate descriptions based on pictures as well as images based on text (\citeauthor{[50]}, \citeyear{[50]}). The LLM focuses on natural language processing tasks and excels at generating, understanding, and engaging in dialogues with textual content in interactions (\citeauthor{[51]}, \citeyear{[51]}). GAN, unlike transformers that utilise self-attention mechanisms to process incoming sequential data, generates highly realistic images through adversarial training and is often used to simulate real-world visual content (\citeauthor{[46]}, \citeyear{[46]}). Another standard model used for image generation is the Diffusion Model, which generates images in a stepwise denoising manner and specialises in the generation of detailed and structured visual content (\citeauthor{[52]}, \citeyear{[52]}).

In XR applications, the diffusion model, particularly the Stable Diffusion Model, is primarily utilised in virtual environments or for generating high-quality details and textures. Compared to the Large Language Model (such as the GPT series), Stable Diffusion excels at generating high-quality images and three-dimensional objects in XR, significantly enriching the visual effects of virtual worlds. Through the conversion of text to images, users can quickly create scenes, people, or objects on demand, thereby enhancing the immersion and interactivity of the XR system. The primary application of Large Language Models in XR is in text interactions and voice conversations, where they are often utilised to provide intelligent assistant functionality, task guidance, or facilitate conversational interactions. GANs in XR can be used to generate highly realistic virtual characters or scenes to help create immersive virtual environments. The visual language model, as a multi-modal input model (accepting image + text), is still in the exploratory stage in the application of XR and has only been used in a few studies. However, it is undeniable that this type of model still holds research potential in integrating multimodal understanding and XR perception. The application of the transformer as a basic model in XR is mainly reflected in the processing of language and text-related tasks. Therefore, the application of different types of generative models in the XR system has its characteristics, which bring unique advantages for creating the virtual world and enhancing user interaction.


\begin{figure}[t]
\centering
\includegraphics[width=1\columnwidth]{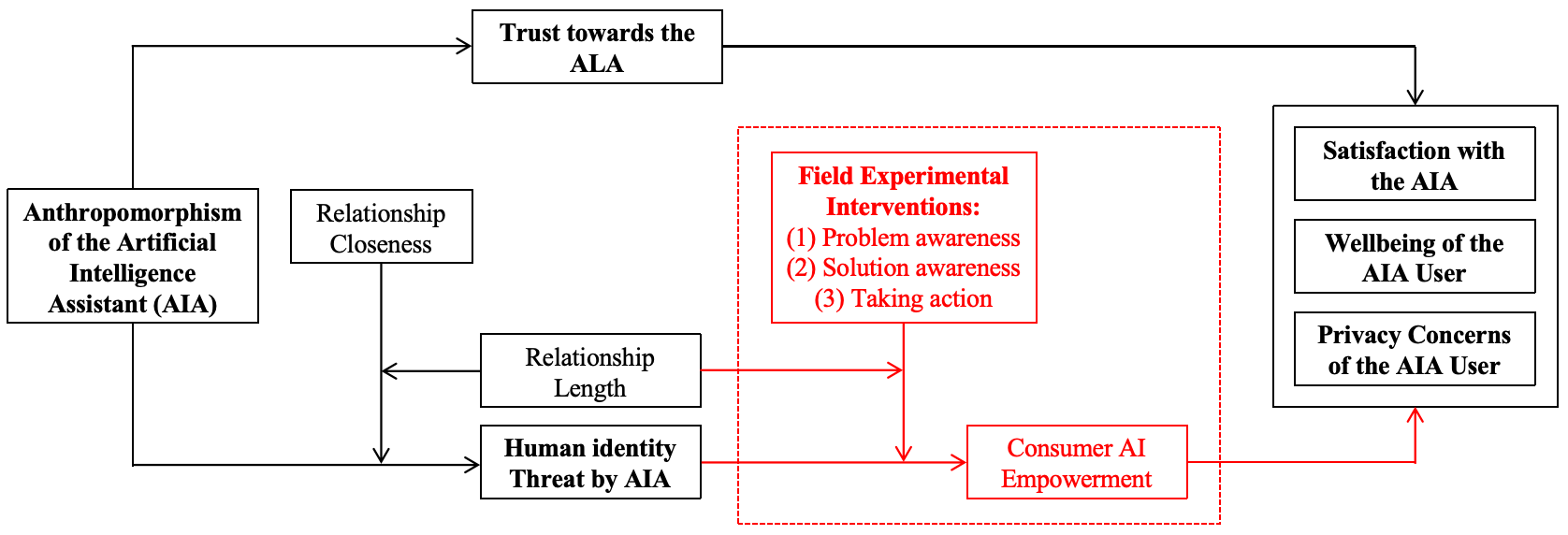}
\caption{A relationship perspective on artificial intelligence assistants with human-like features. }
\label{fig:f8}
\end{figure}

\subsection{The Fusion of Generative AI and XR}
Among these integration approaches, the two most common modes of generative AI application in XR are object generation based on virtual environments and dynamic content generation based on user input.
In the mode of object generation based on virtual environments, generative AI is utilised to create or enhance static content in XR environments, with user input typically being indirect, primarily in the initial settings or preference selection. Once the environment is established, the AI automatically generates virtual objects or scenes based on preset parameters, allowing the user to explore and interact within them. The AIsop system (\citeauthor{[26]}, \citeyear{[26]}), utilises a large language model to automatically generate narrative experiences in virtual reality. The system generates story content from a large language model. It combines text-to-speech technology for narration with diffusion modelling to create visual elements that match the storyline, thus providing a rich and immersive storytelling experience.
In the dynamic content generation mode based on user input, generative AI can respond to specific user inputs (e.g., speech, text, gestures, etc.) in real-time to instantly generate personalised content or adjust the virtual environment (\citeauthor{[53]}, \citeyear{[53]}). User input is the direct trigger for content generation, and this approach emphasises a high degree of interactivity and customisation over virtual environment-based object generation. For example, the Dream Mesh (\citeauthor{[18]}, \citeyear{[18]}) system allows users to describe a desired 3D model by speech, which is converted into text and then generated into a corresponding 3D model that is rendered in a Mixed Reality environment. Building on this, researchers have continued to explore and introduced the Matrix (\citeauthor{[72]}, \citeyear{[72]}), which compared generation time and other factors with Dream Mesh. The results showed that Matrix outperforms Dream Mesh in terms of generation speed and efficiency, providing a more efficient solution for user-driven content generation. In addition, the MS2Mesh-XR pipeline (\citeauthor{[17]}, \citeyear{[17]}) combines hand-drawn sketches and voice input to achieve rapid generation of high-quality 3D mesh models. Users draw sketches in virtual environments through hand gestures supplemented by voice descriptions, and the system generates detailed 3d models for real-time visualisation and manipulation by users \textcolor{black}{with MS2Mesh-XR}. However, in some complex application scenarios, such as operating table simulations (\citeauthor{[54]}, \citeyear{[54]}), the accuracy and realism of the generated content are crucial. Therefore, current generative AI techniques still face challenges in meeting these high standards.

\textcolor{black}{These recent systems illustrate actionable technical pathways for integrating diffusion models, vision-language models, and LLMs in XR content creation. They provide practitioners with concrete examples of how multi-modal generative AI can be architected for both automated and interactive XR experiences. However, it is important to note that in complex application scenarios, such as surgical simulations, the accuracy and realism of generated content remain challenging, highlighting ongoing research needs in this area.}

\subsection{Technical Challenges and Research Gaps}

\textcolor{black}{Currently, generative AI proficiently generates distinct material, including graphics, music, writing, and code. Although Generative AI has shown great potential for application in Extended Reality (XR), there are still many technical challenges to achieving efficient and stable generative interfaces, as well as rich and personalised user experiences, simultaneously (\citeauthor{[32]}, \citeyear{[32]}). Nonetheless, interface design is intrinsically distinct, user behaviour is complicated, and user satisfaction is hard to quantify and explain. It requires comprehension of context, user purpose, emotional subtleties, and the fluid progression of dialogue or action. The next advancement of generative AI in human-computer interaction must focus on not just generating artefacts but also orchestrating user experiences. This entails comprehending the user's objectives, modifying interfaces instantaneously, and even predicting requirements. The objective is to transition from static outputs to dynamic, context-sensitive systems capable of co-designing and co-evolving with users. \\ 
A paradigm change in the study of human-computer interaction will occur in the new era of generative AI. Human users, both in the past and now, acquire skills to navigate interfaces; nevertheless, interfaces powered by generative AIs will adapt to interact with people. }
The following will summarise the major technical bottlenecks and research gaps in the current field based on 26 papers.

\subsubsection{Deep Integration of Multi-Modal Interaction}
\textcolor{black}{Recent advances in real-time, context-aware 3D model generation within XR environments have been significantly propelled by the integration of vision-language models and multi-modal user inputs, such as speech and gesture. Several recent contributions have demonstrated the feasibility and effectiveness of leveraging these technologies to enable more natural and intuitive interactions in XR (  \citeauthor{[73]}, \citeyear{[73]}; \citeauthor{[74]}, \citeyear{[74]}; \citeauthor{[72]}, \citeyear{[72]}). For example, systems that combine visual scene understanding with natural language processing can dynamically generate or modify 3D content in response to spoken commands or gestural cues, allowing users to interact with virtual environments in a manner that closely mirrors real-world communication. These developments exemplify the deep integration of multi-modal interaction, as they not only enhance user experience but also expand the creative and collaborative possibilities within XR. Incorporating such illustrative examples would strengthen the discussion of multi-modal interaction, highlighting the practical impact and ongoing innovation in this rapidly evolving field. }

Although studies have explored multi-modal interactions such as speech-to-3D model generation, how to realise the deep integration between different modalities is still an urgent issue. The challenge of multimodal interaction lies not only in how to effectively combine inputs from visual, auditory, tactile, and other modalities, but also in how to design adaptive interactions that enhance the user experience. Generative AI technologies play a central role in this process. 
\textcolor{black}{Significantly, human-computer interaction is intrinsically multimodal and often embodied, as shown by augmented reality, virtual reality, robots, and intelligent surroundings. Generative AI must acquire the ability to synchronise between modalities and physical settings, creating not only screens but experiences that include both digital and physical realms.} 
For example, one study (\citeauthor{[32]}, \citeyear{[32]}) explores the application of generative AI in the metaverse, emphasising the convergence of content generation and transmission. Selected articles (\citeauthor{[18]}, \citeyear{[18]}; \citeauthor{[23]}, \citeyear{[23]}) involve multi-modal interactions, but further research is needed to realise deep integration and adaptive interactions between different modalities. With more accurate algorithms and enhanced system architectures, future research needs to address this deep integration problem to improve the naturalness and responsiveness of the interaction. 

\subsubsection{Latency and Stuttering}
Generating models, especially large language models and image generation models, typically requires substantial computational resources. However, in XR applications, where real-time is critical, generative models need to be deployed to minimise latency and improve responsiveness. Current research focuses on deploying generative models to edge devices to reduce transmission latency and enhance real-time performance (\citeauthor{[55]}, \citeyear{[55]}). A multi-modal sketch-to-mesh generation approach in MS2Mesh highlights the challenges of edge computing in terms of real-time and computational requirements (\citeauthor{[17]}, \citeyear{[17]}). The optimisation of model compression, quantisation, and refinement techniques to efficiently implement generative models on edge devices still needs further exploration. With the gradual increase in realism and latency-free requirements for virtual worlds, the computational demand continues to rise, and improving the efficiency of edge computing without sacrificing performance will become a key future direction to explore (\citeauthor{[56]}, \citeyear{[56]}).

\subsubsection{System Integration and Standardisation}
The application of generative AI in XR usually requires integration with existing systems. However, there is a lack of unified standards and interfaces, which limits the broad application of the technology and the interoperability between different systems. With the continuous development of XR technology, developing unified standards and designing modular and extensible system architectures has become a key issue to be addressed. Although some articles have proposed the integrated application of generative AI in XR systems, they fail to provide a comprehensive standardised solution (\citeauthor{[15]}, \citeyear{[15]}; \citeauthor{[12]}, \citeyear{[12]}). Therefore, to promote the diffusion and popularisation of generative AI technology, future research can focus on innovations in system architecture, data transmission interface design, and other areas to enhance the system's interoperability and scalability.

\subsubsection{AI-enabled virtuality in physical worlds}
\textcolor{black}{While this survey article notes that virtual reality (VR) is more extensively represented in the existing literature, it is equally important to draw attention to the rapid progress being made in augmented reality (AR)-specific generative AI pipelines (\citeauthor{[72]}, \citeyear{[72]}). Recent research has demonstrated that AR environments, when combined with generative AI, can enable in-situ, user-driven object creation and manipulation, fundamentally transforming how users interact with digital content in real-world contexts ( \citeauthor{[73]}, \citeyear{[73]}; \citeauthor{[45]}, \citeyear{[45]}; \citeauthor{[55]}, \citeyear{[55]}). These advances are particularly significant for application domains such as design, education, and remote collaboration.
\\
For example, in design, generative AI models integrated with AR platforms allow users to visualize, generate, and iteratively refine 3D models directly within their physical surroundings (\citeauthor{[74]}, \citeyear{[74]}). This not only streamlines the creative process but also fosters a more intuitive and immersive workflow, as designers can immediately assess how virtual objects interact with real-world constraints (\citeauthor{[40]}, \citeyear{[40]}). In education, AR-based generative AI tools can facilitate interactive and personalized learning experiences, enabling students to create, manipulate, and explore virtual objects or scientific phenomena in real time, thereby enhancing engagement and comprehension  (\citeauthor{[25]}, \citeyear{[25]}).
\\
Remote collaboration also stands to benefit significantly from these developments. AR systems powered by generative AI can support shared, context-aware object creation and manipulation, allowing geographically distributed teams to co-design, annotate, and modify virtual content as if they were physically present together. Such capabilities can improve communication, reduce misunderstandings, and accelerate decision-making processes (\citeauthor{[53]}, \citeyear{[53]}). 
By highlighting these AR-specific generative AI pipelines, the review would provide a more comprehensive and balanced perspective, showcasing how these technologies are poised to augment real-world experiences and drive innovation across multiple sectors. This focus would also underscore the growing importance of AR in the broader landscape of generative AI research and applications. }



\section{CHALLENGES}
\subsection{Security and Privacy Challenges}
In an extended reality (XR) environment, the system collects a wide range of user data, including biometric information such as eye tracking and facial recognition data, as well as the user’s personal habits and preferences (\citeauthor{[57]}, \citeyear{[57]}). This data is a vital resource for improving user experience and training generative AI models.
However, this data collection also presents significant challenges to privacy and security. In the event of a data breach, not only will users’ Personally Identifiable Information (PII) be exposed, but also sensitive details such as their behaviour patterns and usage habits may become public (\citeauthor{[58]}, \citeyear{[58]}). Such breaches provide a wealth of material for cybercriminals to execute more precise phishing and social engineering attacks. For example, hackers may use the leaked data to construct highly personalised fraudulent communications, trick users into providing more personal information, or directly induce them to conduct financial transactions, resulting in severe financial losses and personal privacy violations.

To effectively prevent security incidents, we must deploy comprehensive and robust encryption measures to ensure that data is adequately protected both during transit and in storage (\citeauthor{[59]}, \citeyear{[59]}). At the same time, we should follow the principle of “data minimisation” and collect only the necessary information that is essential to the realisation of our AI capabilities. This approach can not only help protect users’ privacy but also reduce possible legal exposure.
According to Ometov’s survey (\citeauthor{[60]}, \citeyear{[60]}), Multi-Factor Authentication (MFA) is a powerful tool for improving authentication security. In addition to traditional password authentication, biometrics such as fingerprint recognition and iris scanning can be combined with other authentication methods, including dynamic passwords or hardware keys (\citeauthor{[61]}, \citeyear{[61]}). Such a combination can not only significantly improve the security of the system, but also effectively resist malicious attacks, providing users with a more robust security barrier.

When we combine AI with XR technology, ensuring data security and personal privacy for every user becomes critical, which is the cornerstone of long-term growth. We all strive for the perfect balance between innovation and security (\citeauthor{[62]}, \citeyear{[62]}) - on the one hand, innovation is an indefatigable explorer, constantly pushing technology forward and opening doors to new experiences; on the other hand, security is like a loyal guard guarding these gates, ensuring that every friend who steps through them can enjoy the journey without worrying about the shadow behind them.

\subsection{User Autonomy-decision-ability Challenge}
As the convergence of artificial intelligence (AI) and Extended Reality (XR) technologies grows, so does public concern about the autonomy and decision-making capabilities of these systems (\citeauthor{[63]}, \citeyear{[63]}). As such systems become more complex and powerful, users may increasingly rely on them for cognitive processing and decision support. There is a potential risk that the user’s ability to think independently and solve problems may be inadvertently compromised. Due to the convenience of AI systems, users are becoming accustomed to outsourcing complex cognitive tasks to AI, which may lead to a degradation of some of their cognitive skills over time (\citeauthor{[64]}, \citeyear{[64]}; \citeauthor{[65]}, \citeyear{[65]}). Figure 7 illustrates the relationship between artificial intelligence assistants with human-like features.

As a result, a central question arises: how much cognitive control should we grant to AI in an extended reality (XR) environment? At what point does this dependency change from an aid to a stumbling block to our cognitive growth? When people over-rely on AI to handle complex thinking tasks, it can lead to a decline in creativity and critical thinking skills, as they no longer actively explore new knowledge or analyse problems in depth. In the workplace, if employees are accustomed to letting AI complete daily tasks, they may miss the opportunity to exercise their judgment and thus lack the ability to respond to emergencies.

To address these issues, developers should aim to design generative AI systems that provide effective support without overly interfering with user autonomy. For example, in XR education scenarios, AI can play the role of mentor while inspiring students to think independently and discover themselves. Additionally, there is a need to disseminate knowledge about how to effectively utilise AI tools, particularly among young people, whose critical thinking skills are of utmost importance. The goal is to ensure that people can enjoy the convenience of technology while still maintaining and developing their cognitive abilities and problem-solving skills (\citeauthor{[64]}, \citeyear{[64]}). In this way, we can promote the development of a social environment that values both technological innovation and human intelligence.

The combination of AI and XR offers endless possibilities for the future. However, to maximise the positive impact of generative AI on XR and minimise potential risks, we must carefully consider how we allocate cognitive control over AI, ensuring it can complement rather than hinder human cognitive development. In the future, a suitable strategy must be developed to create a positive environment that fosters human growth and technological progress.

\subsection{Child Protection Concerns}
In an extended reality (XR) environment, if AI-generated content contains illegal or inappropriate information, such as violent and bloody scenes, this will not only negatively impact the user’s experience but may also affect their real-life behaviour patterns (\citeauthor{[66]}, \citeyear{[66]}). Especially for children whose values are not yet fully formed, exposure to such content may induce an imitation effect, in which they attempt to imitate the behaviour they see, which can be potentially harmful to their development and social behaviour (\citeauthor{[67]}, \citeyear{[67]}).
In addition, the phenomenon of “desensitisation” refers to frequent exposure to violent or gory content, which may lead to a dulling of an individual’s emotional response to these acts, thereby reducing sensitivity to real-life violence. This change in mental state may further affect a person’s moral judgment and mental health (\citeauthor{[68]}, \citeyear{[68]}). The lack of regulation of AI-generated content could exacerbate children’s psychological problems and increase their acceptance of bad behaviour. 

In the future, we must pay special attention to the algorithmic bias and inappropriate content (\citeauthor{[69]}, \citeyear{[69]}) that may be introduced by generative AI, and we should conduct regular evaluations of XR systems embedded with generative AI, including through penetration testing and code reviews to identify potential vulnerabilities and security risks in XR systems, especially those related to AI-generated content. We should also strengthen the censorship mechanism for AI-generated content to ensure that all content containing violence, pornography, or other harmful information is filtered out, thereby preventing generative AI from producing inappropriate material.
At the same time, the introduction of parental control functions in the XR environment is one direction that can be studied in the future, enabling parents to coordinate the supervision and management of their children’s use and help them avoid inappropriate content. Creating a safer and more nurturing XR environment for children, one that promotes their healthy growth and is free from harmful information, will not only help protect them from potential psychological harm but also support their development of the right values in the digital world, thus effectively safeguarding their rights (\citeauthor{[70]}, \citeyear{[70]}).
The way forward should be to promote cross-platform cooperation. In the face of the challenges posed by XR technology, the strength of any one company or organisation is not enough. We need to strengthen cross-platform collaboration and encourage sharing between XR vendors and industry partners to jointly develop integrated solutions.

\subsection{Long-term Performance \& Sustainability}
Most of the existing research has focused on short-term evaluation in laboratory Settings, and there is a lack of long-term performance and sustainability testing of generative AI and XR combined systems in real-world Settings. In practical application scenarios, the complexity of the environment and external factors may affect the stability and efficiency of the system. 
The use of generative AI in XR should be validated over time, especially in scenarios where continuous operations are required (\citeauthor{[46]}, \citeyear{[46]}). For example, in the medical and industrial sectors, where robots are often required to perform repetitive tasks, research should evaluate the durability, maintainability, and long-term user experience of systems that combine generative AI and XR technology. Future research should focus on testing the performance of XR systems in different real-world environments. For example, the operating environment in industrial applications is more complex, and the system must consider additional external factors, such as equipment failure and environmental interference. Ensuring the stability and efficiency of generative AI applications in complex XR environments is a direction worthy of further study.

\subsection{AI Inspires the Potential of XR in Remote Collaboration}
Although XR technology provides a unique application platform for generative AI, its potential in remote collaboration is still not fully realised. Current research focuses more on the basic applications of XR in enhancing user interactions (\citeauthor{[78]}, \citeyear{[78]}; \citeauthor{[79]}, \citeyear{[79]}), while relatively little research has been conducted on how to enhance interaction effects through generative AI further. 
\textcolor{black}{Generative AI can create real-time, shared collaboration spaces where distributed participants engage as if co-located, enhancing co-presence, interactivity, and engagement (\citeauthor{[71]}, \citeyear{[71]}; \citeauthor{[72]}, \citeyear{[72]}). Recent systems exemplify this potential. BlendScape enables end-users to customize video-conferencing environments by blending physical or virtual backgrounds into AI-generated scenes (\citeauthor{[80]}, \citeyear{[80]}). Such dynamic environments support diverse activities, from design brainstorming to creative workshops, and help alleviate common issues such as reduced engagement and meeting fatigue
. SpaceBlender extends this concept to immersive VR by blending multiple users’ physical surroundings into unified 3D spaces (\citeauthor{[81]}, \citeyear{[81]}). Unlike conventional synthetic VR, these context-rich environments preserve familiar cues, supporting deixis, mutual awareness, and collaboration. A user study reported that participants experienced improved comfort, navigability, and task coordination when working within SpaceBlender-generated environments. Beyond environment design, AI’s data processing capabilities enable the analysis of large datasets and their presentation as intuitive visualisations within XR environments. In remote collaboration, such shared visual spaces allow distributed teams to monitor project status, adapt strategies in real time, and coordinate more effectively. AI can also function as an intelligent assistant in XR by retrieving information, generating reports, or producing design elements through simple voice commands, thereby streamlining routine tasks and improving overall collaboration efficiency.} 


\subsection{\textcolor{black}{Challenges of AI and Social XR}}

\textcolor{black}{In addition to the challenges discussed above, Social XR introduces unique concerns. Social XR refers to multi-user extended reality environments that support shared presence and interaction, such as virtual meetings, collaborative workspaces, and online social gatherings. These environments are becoming increasingly important for remote communication and may play a central role in future digital society by augmenting how people connect, collaborate, and build trust across distance. }

\textcolor{black}{The integration of AI into Social XR creates new challenges beyond those of single-user XR. On the technical side, issues such as real-time responsiveness, multi-user synchronization, and scalable AI-driven content generation remain difficult, as even small delays or inconsistencies can disrupt social presence (\citeauthor{[82]}, \citeyear{[82]}). On the ethical side, identity authentication and protection against deepfakes are critical, since AI can generate highly realistic avatars or voices that may be misused for impersonation. Similarly, calibrating trust in AI agents is essential to avoid overreliance or deception. On the user experience side, maintaining social presence and emotional realism is challenging: mismatches in avatar appearance, speech, or behavior risk triggering discomfort and eroding engagement. Addressing these technical, ethical, and experiential challenges is essential for ensuring that AI-enhanced Social XR fosters immersive, trustworthy, and meaningful shared experiences rather than undermining them.}

\subsection{\textcolor{black}{Compliance and Legal Challenges in AI and XR}}
\textcolor{black}{As AI increasingly underpins XR and metaverse platforms, compliance and legal risk shift from isolated data protection to system-level governance. GenAI and LLM amplify expressive power and personalization, but compound bias, misinformation, and cross-context data flows, demanding customized ethical legal frameworks and accountability regimes (\citeauthor{[83]}, \citeyear{[83]}). Transparency is a prerequisite for compliance. XR-specific XAI is needed to make XR-specific XAI intelligible to users, auditors, and regulators, and to calibrate trust without eroding immersion; current evidence shows a gap in XR-adapted methods and evaluation practice (\citeauthor{[84]}, \citeyear{[84]}). Security and privacy are also critical to safety in multimodal, real-time AI-XR: adversarial manipulation of models, identities, and streams can degrade safety, violate confidentiality, and corrode trust; emerging taxonomies point to defenses such as robust training, authenticated avatars, privacy-preserving analytics, and continuous monitoring, but coverage remains incomplete (\citeauthor{[85]}, \citeyear{[85]}). Beyond doctrinal compliance, AI-XR poses a deeper jurisprudential challenge: by modulating perception and prediction, systems can steer desires and choices, effectively 'engineering regulatees' in ways that evade traditional tests for coercion or undue influence, making harms difficult to name or redress (\citeauthor{[86]}, \citeyear{[86]}). 
Therefore, developing a responsible AI-XR requires integrated transparency, robustness, and design that preserves rights, coupled with updated legal concepts of agency, manipulation, and accountability.}

In summary, by combining advanced AI technology with immersive XR experiences, we not only enhance the efficiency and effectiveness of remote collaboration but also introduce innovative changes to education and training, promoting the development of personalised learning. This combination not only increases productivity but also guarantees quality learning and development opportunities for all participants. Remote collaboration can not only become more efficient and flexible but also increase interaction and engagement between team members. This combination opens up many innovative possibilities for future work.

\section{CONCLUSION}
This survey article examines the application of generative AI in conjunction with Extended Reality (XR) technology by systematically reviewing and analysing 26 related studies published within the last three years. Though a small number of articles are being collected, the purpose of this study is to provide academia and industry with an understanding of the potential, challenges, and future research directions of Generative AI in XR. By adopting the PRISMA methodology, this study gives a first look into the technical implementations of generative AI in various application areas of XR technology, encompassing a broad range of fields such as design, education, and training, and reveals its significant potential for enhancing user experience, interactivity, and creativity.

It is found that the application of generative AI in XR is redefining the way of content generation and user interaction in virtual environments, especially in Augmented Reality (AR) and Virtual Reality (VR) environments, which significantly enriches the immersiveness and interactivity of XR systems through techniques such as Natural Language Processing, Image Generation and Action Generation. However, although generative AI demonstrates great potential for application in XR, current research still faces many challenges, especially in the areas of multi-modal interaction, latency, real-time system integration and standardisation. To this end, this study proposes that future research should focus on enhancing the deep integration of generative AI and XR technologies, promoting user-friendliness and diversity in system design, and meeting growing technological demands and application scenarios. In addition, ethical issues such as security and privacy protection, safeguarding users’ autonomous decision-making ability, and child protection are also important topics that will need to be addressed in the future.

Overall, this \textcolor{black}{concise review article} offers valuable insights and guidance on the integration of generative AI and XR technologies, highlighting the trends, challenges, and potential opportunities presented by current technology applications. With the continuous advancement of generative AI technology and the development of XR systems, this cross-cutting area is poised to become a significant frontier for technological innovation in the future. We call for additional research efforts by presenting this study, which intends to provide valuable insights for academic researchers, technology developers, and practitioners in related industries, promoting in-depth research and exploration of generative AI in XR and facilitating the development of more innovative solutions.

\section*{Acknowledgement(s)}


\textcolor{black}{This research was partially supported by the Hong Kong Polytechnic University's Start-up Fund for New Recruits (No. P0046056), Departmental General Research Fund (DGRF) from HK PolyU ISE (No. P0056354), and PolyU RIAM -- Research Institute for Advanced Manufacturing (No. P0056767). Xian Wang was supported by a grant from the PolyU Research Committee under student account codes RMHD.}

\bibliographystyle{apalike2}
\bibliography{bibtex}

\bigskip

\section*{About the Authors}

\begin{description}
\item[Xinyu Ning] received an M.Sc. in Industrial and Systems Engineering from the Hong Kong Polytechnic University, Hong Kong S.A.R. She also holds a BEng degree in Mechanical Engineering and English. Her research interests primarily focus on human-computer interaction, specifically combining mixed reality technologies.
\item[Zhuo Yan] is a graduate student in the Smart Manufacturing program at the Hong Kong Polytechnic University. Her research interests include smart manufacturing systems, artificial intelligence, and human-computer interaction technologies. Zhuo Yan aims to drive innovation in smart manufacturing and promote sustainable development and intelligent transformation through her research.
\item[Xian Wang] received M.Phil. degree in Artificial Intelligence at the Hong Kong University of Science and Technology. She is currently a Ph.D. student at the Hong Kong Polytechnic University. Her research interests include human-computer interaction, haptic feedback, virtual reality, and collaboration.
\item[Chan-In Devin SIO] is a Ph.D. candidate in the Department of Industrial and Systems Engineering at The Hong Kong Polytechnic University. His research interests include Metaverse Humanity, human–computer interaction, and technology-mediated legal processes. He also holds a postgraduate qualification in Sustainability Leadership and professional legal credentials.
\item[Lik-Hang Lee] is currently an assistant professor with the Hong Kong Polytechnic University. He received a Ph.D. degree from the Hong Kong University of Science and Technology, and Bachelor's and M.Phil. degrees from the University of Hong Kong. His research interests are augmented and virtual realities (AR/VR).
\end{description}

\end{document}